\documentclass{INTERSPEECH2023}

 \interspeechcameraready

\usepackage{multirow, tabularx}
\usepackage{booktabs}
\usepackage{xcolor}
\usepackage{amssymb}
\usepackage{enumitem}
\setitemize{noitemsep,topsep=0pt,parsep=0pt,partopsep=0pt}
\usepackage{algorithm}
\usepackage{algpseudocode}

\title{Deep Learning for Joint Acoustic Echo and Acoustic Howling Suppression in Hybrid Meetings}
\name{First Author Name$^1$, Second Author Name$^2$, Third Author Name$^3$}
\name{Hao Zhang, Meng Yu, Dong Yu}
\address{Tencent AI Lab, Bellevue, WA, USA}
\email{ \{aaronhzhang, raymondmyu, dyu\}@global.tencent.com}

\begin{document}

\maketitle
 
\begin{abstract}
Hybrid meetings have become increasingly necessary during the post-COVID period and also brought new challenges for solving audio-related problems. 
In particular, the interplay between acoustic echo and acoustic howling in a hybrid meeting makes the joint suppression of them difficult. 
This paper proposes a deep learning approach to tackle this problem by formulating a recurrent feedback suppression process as an instantaneous speech separation task using the teacher-forced training strategy. Specifically, a self-attentive recurrent neural network is utilized to extract the target speech from microphone recordings with accessible and learned reference signals, thus suppressing acoustic echo and acoustic howling simultaneously. 
Different combinations of input signals and loss functions have been investigated for performance improvement. 
Experimental results demonstrate the effectiveness of the proposed method for suppressing echo and howling jointly in hybrid meetings.

\end{abstract}
\noindent\textbf{Index Terms}: hybrid meetings, acoustic echo cancellation, acoustic howling suppression, teacher forcing training

\section{Introduction}



Hybrid meetings, which involve a combination of in-person and remote participants, have become increasingly essential in the post-COVID era \cite{saatcci2019hybrid, hameed2021will}. As of 2022, a significant proportion of workplaces ($78\%$) have adopted hybrid work strategies, indicating a growing trend towards hybrid work as the future of work \cite{WinNT}.
However, despite the benefits of hybrid meetings, audio-related problems such as acoustic echo and acoustic howling can pose significant challenges and need to be addressed to ensure full-duplex communication.



Acoustic echo refers to the phenomenon where sound originating from a speaker on one end of a communication system is captured by the microphone on the other end and subsequently replayed back to the speaker, creating an unwanted echoing effect \cite{benesty2001advances, enzner2014acoustic}. 
Acoustic howling arises when sound from the speaker's end is captured by the microphone on the same end, leading to a feedback loop that amplifies the sound until it becomes unbearable \cite{waterhouse1965theory, van2010fifty}. 
Despite having similar underlying mechanisms, acoustic echo and howling are distinct problems, and they can be particularly challenging to address in hybrid meetings where both issues can occur simultaneously.
Therefore, it is crucial to have robust and effective algorithms that can address both acoustic echo cancellation (AEC) and acoustic howling suppression (AHS) in a joint manner, taking into account the complex acoustics of the hybrid meeting environment. However, the presence of one problem can affect the estimation and suppression of the other, making it difficult for conventional algorithms to effectively suppress both echo and howling jointly.


Recently, deep learning has emerged as a promising approach for solving the challenges of AEC and AHS due to its ability to model complex nonlinear relationships \cite{zhang2018deep, zhang2022deep, zhang2023deep, chen2022neural, gan2022howling, zheng2022deep, zhang2023deepAHS}. 
In AEC, the problem can be directly formulated as a supervised speech separation problem \cite{lee2015dnn, zhang2018deep, zhang2019deep}. However, AHS poses a more complex challenge since it involves the recursively amplification of the playback signal, which makes formulating it as a supervised learning problem non-trivial.
To address this challenge, Zhang et al. \cite{zhang2023deepAHS} recently proposed a deep learning based AHS method (Deep AHS) using teacher-forced training strategy, resulting in improved performance when compared to baselines.
We believe that recent advances in deep learning based AEC and AHS make it possible to develop effective deep learning methods to address them jointly and solve the full-duplex communication problem in hybrid meetings. 



%
In this study, we tackle the challenges posed by joint AEC and AHS by considering them as an integrated feedback suppression problem and propose a deep learning approach to address it. 
The recursive feedback suppression process is converted to a speech separation process through teacher forcing training strategy \cite{williams1989learning, lamb2016professor}, which simplifies the problem formulation and accelerates model training. 
To accomplish this task, a self-attentive recurrent neural network (SARNN) \cite{yu2022neuralecho} is utilized to extract target speech from microphone signal with multiple reference signals as additional inputs. 
Various combinations of inputs are explored to take full use of the accessible reference signals.
Given the difficulties in suppressing both forms of feedback jointly, a specific loss function is designed to mitigate leakage introduced due to improperly suppressed feedback, with results demonstrating its efficacy.
Experimental results show the effectiveness of the proposed method for joint echo and howling suppression.
%

%

The structure of this paper is as follows: Section 2 provides an overview of the audio-related issues in hybrid meeting systems. Section 3 presents the proposed method. The experimental setup is outlined in Section 4, and Section 5 reports the corresponding results. Section 6 concludes the paper.

\section{Hybrid meetings}


\begin{figure}[!t]
\centering
     \includegraphics[width=0.999\columnwidth]{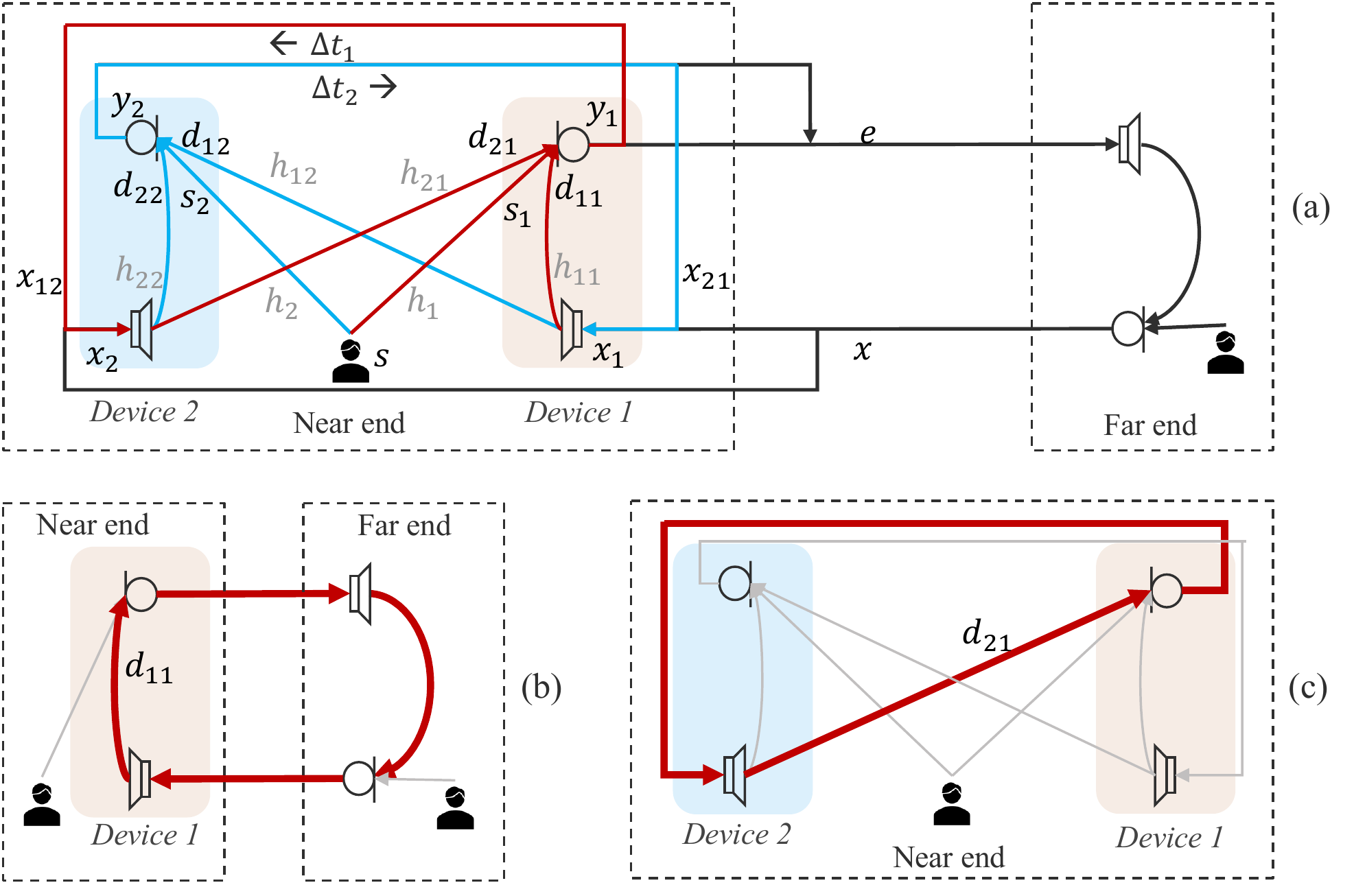}
      \caption{(a) A simplified hybrid meeting system. (b) and (c) illustrate the two closed acoustic loops related to device 1.}
      \label{fig:HybridMeeting}
\end{figure}

\subsection{Signal model in a hybrid meeting system}


For a hybrid meeting system with $J$ devices on the same end and all of them have both a loudspeaker and a microphone turned on, then the total number of acoustic paths in the system will be $J^2$. Take a simplified system with two devices on the same end as an example, as shown in Figure~\ref{fig:HybridMeeting} (a). While capturing the target speech $s_i$, the microphone on device $i$ will also record the background noise $n_i$, and playback signals from all devices: 
\begin{flalign}
\label{equ:mic1}
y_i = s_i + n_i + \sum_{j=1} ^ 2 d_{ji} =  s_i + n_i + \sum_{j=1} ^ 2 (x_j * h_{ji}) 
\end{flalign}
where $x_j$ is the loudspeaker signal on device $j$, and $d_{ji}$ is the signal picked up by microphone $i$ from loudspeaker $j$ through the acoustic path $h_{ji}$. 
Among these playback signals, $d_{ii}$ is the playback from device $i$'s own loudspeaker to its microphone, which is known as acoustic echo. Compared to $d_{ji}$ ($j\neq i$), acoustic echo ($d_{ii}$) is relatively easier to suppress since each device usually only has access to its own loudspeaker signal $x_i$, which can be used as a reference signal during the attenuation of $d_{ii}$. 

Challenges arise when speakers on the far end and near end talk simultaneously. 
Considering that each device cannot distinguish whether other devices are exposed in the same space or not, it treats all other devices as far end and sends its processed signal to them. 
The loudspeaker signal $x_i$ will then be a combination of the far-end signal $x$ and the processed signals sent to device $i$ from device $j$ (denoted as $x_{ji}, (j\neq i)$):
\begin{flalign}
\label{equ:ref}
x_i =x + x_{ji} , j\neq i
\end{flalign}
%
%
%
%
If feedback suppression module on each device works properly, the resulting processed signal, $x_{ji}$, should resemble a delayed, scaled, and reverberant version of the near end speech $s$. From the perspective of signal sources, microphone signal given in (\ref{equ:mic1}) can be rewritten as:
\begin{flalign}
\label{equ:playback}
y_i = s_i + n_i + \sum_{j=1} ^ 2 d_{ji}^x +\sum_{j=1} ^ 2  d_{ji}^s
\end{flalign}
where $d_{ji}^x$ and $d_{ji}^s$ represent the playback components originated from $x$ and $s$, respectively. 
It is more challenging to suppress $d_{ji}^s$ because it comes from the same source as that of the target speech $s_i$, and reducing it could distort the target signal.

\subsection{Joint acoustic echo and acoustic howling suppression}
%



Let us focus on device 1 to analysis the audio-related problem in a hybrid meeting system. 
There are two closed acoustic loops (CAL) per device in the system, shown in Figure~\ref{fig:HybridMeeting} (b) and (c), that can cause acoustic howling. 
The first CAL, due to acoustic echo, is easier to handle since acoustic echo occurs once per transmission and is handled on both ends. The second CAL is more challenging due to two reasons: 1) Device $1$ lacks access to the reference signal that causes feedback $d_{21}$. 2) The two devices involved in this CAL are exposed in the same space.

Without any processing, the microphone signal will be played out through loudspeaker and repeatedly re-enter the pickup. The microphone signal $y_1$ at time index $t$ can then be represented as:
\begin{flalign}
\label{equ:howling}
\textstyle y_1(t) = &  s_1(t) + n_1(t) +  d_{11}(t) + \\ \nonumber
& NL\left[(y_1(t-\Delta t_1) + x(t)) \cdot G_2\right]*h_{21}(t) 
\end{flalign}
where $\Delta t_1$ denotes the system delay from device 1 to device 2, $G_2$ is the gain of amplifier on device 2, and $NL(\cdot)$ the nonlinear function of loudspeaker. 
Playback $d_{11}(t)$ is the acoustic echo. 
The recursive relationship between $y_1(t)$ and $y_1(t-\Delta t)$ causes re-amplifying of playback signal and leads to an annoying, high-pitched sound, which is known as acoustic howling. 




In hybrid meetings, achieving full-duplex communication requires addressing both AEC and AHS simultaneously. 
Nonetheless, the presence of either issue can hinder the accurate detection and elimination of the other, resulting in a shortage of effective solutions.


%
%

\section{Proposed method}

\begin{figure}[!t]
\centering
     \includegraphics[width=0.85\columnwidth]{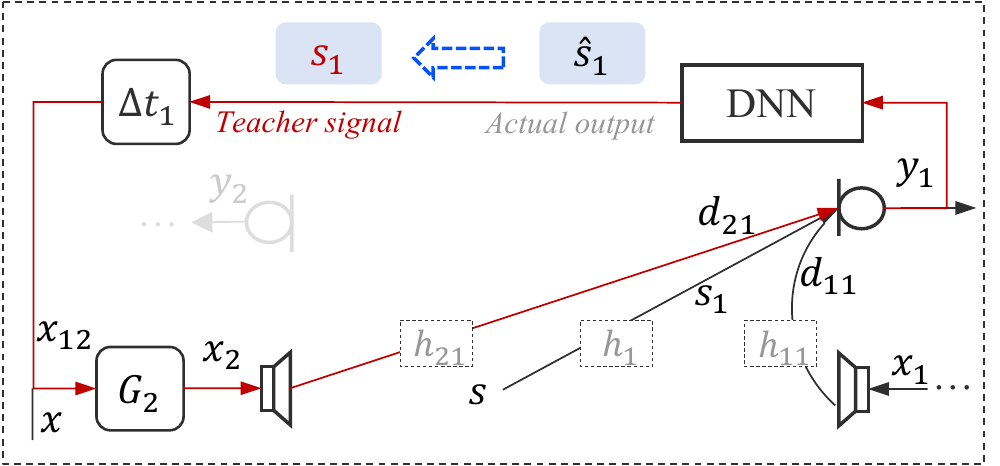}
      \caption{Signal flow with teacher-forced learning strategy.}
      \label{fig:DeepAFS}
\end{figure}

\subsection{Problem formulation}

%


%
%

To address the recursive nature of howling, a deep neural network (DNN) module needs to be integrated into the closed acoustic loop and trained recursively. However, this is not practical due to its high computational cost. 
Alternatively, teacher forcing training strategy can be used to formulate the joint AEC and AHS task as a general feedback suppression problem, as is detailed in Figure~\ref{fig:DeepAFS}. This is based on the assumption that the model, once properly trained, can attenuate all feedback signals ($d_{11}$ and $d_{21}$) and transmit only the target speech $s_1$. 
Through teacher-forced learning, the actual output $\hat{s}_1$ is replaced with the teacher signal $s_1$ during model training. 
As a result, rather than generated recursively, the microphone signal (\ref{equ:howling}) is simplified to a mixture of target signal, background noise, acoustic echo, and an one-time playback signal determined by $s_1$:
\begin{flalign}
\label{equ:mic}
y_1(t)  \textstyle = &s_1(t) + n_1(t)+ d_{11}(t) + \\ \nonumber
&NL[(s_1(t-\Delta t_1)+x(t)) \cdot G_2] *h_{21}(t)
\end{flalign}
And the overall problem can thus be formulated as a speech separation problem during model training where the task is to separate target signal from the microphone recording with accessible loudspeaker signals ($x_1$, and/or $x, x_{21}$) as references.



\subsection{Inputs and reference signals}

%

Appropriate reference signals, which enables accurate estimation of the playback signals, are crucial for AEC and AHS algorithms.
The reference signal for device 1 is a mixture of two signals, as shown in Figure~\ref{fig:HybridMeeting} and (\ref{equ:ref}). The most direct approach is to use the integrated signal $x_1$ as a reference for suppressing the two feedback signals $d_{11}$ and $d_{21}$ in $y_1$. However, this may be less effective for suppressing $d_{21}$. 
Known from equation (\ref{equ:playback}) that the playback signals share common components originating from different sound sources. 
Depending on the design of the audio system, we could also have access to $x$ and $x_{21}$ in addition to the integrated loudspeaker signal $x_1$. 
Using separated loudspeaker signals ($x$ and $x_{21}$) as references could make the suppression of both feedbacks more efficient. 

Besides these accessible reference signals obtained directly from device, we have also designed the network to estimate some intermediate outputs from the inputs and use them as nonlinear reference signals to further improve feedback cancellation performance \cite{zhang2022deep, zhang2023kalmannet}. 





\subsection{Network structure}

%
%

\begin{figure}[!t]
\centering
     \includegraphics[width=0.98\columnwidth]{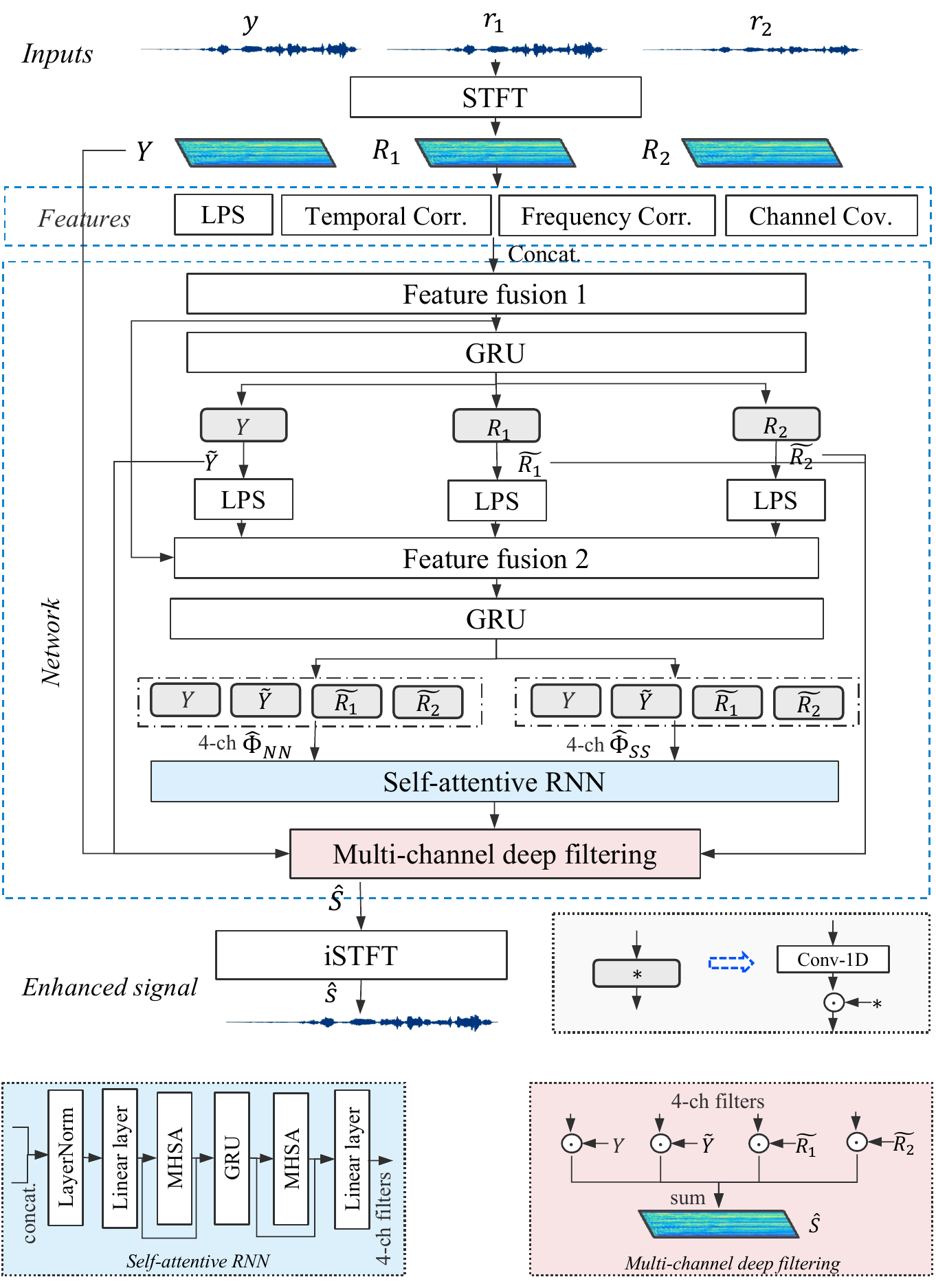}
      \caption{Architecture of the DNN module for joint acoustic echo and howling suppression. Where the block in gray denotes a combination of 1D convolution layer and deep filtering. A ``Conv-1D'' outputs a complex-valued ratio filter, which is then applied upon signal $*$ through deep filtering, denoted as $\odot$. }
      \label{fig:NeuralEcho}
\end{figure}

Network of the proposed method is given in Figure~\ref{fig:NeuralEcho}. It takes the microphone signal and one or two reference signals (represented as $r_1$ and $r_2$) as inputs. The input signals, sampled at 16 kHz, are transformed into the frequency domain using a 512-point short-time Fourier transform (STFT) with a frame size of 32 ms and frame shift of 16 ms. The resulting frequency domain inputs are labeled as $Y$, $R_1$, and $R_2$, respectively. 

To extract more information from inputs and facilitate the suppression of playback signals, we follow \cite{yu2022neuralecho, zhang2023deepAHS} and design the input feature as a concatenation of the normalized log-power spectra (LPS), correlation matrix across time frames and frequency bins, and channel covariance of input signals. These features are concatenated and then passed through a linear layer for feature fusion, followed by a gated recurrent unit (GRU) layer with 257 hidden units and three 1D convolution layers to estimate three complex-valued filters. The filters are then applied to the inputs through deep filtering \cite{mack2019deep} to obtain the corresponding intermediate signals, $\tilde{Y}$, $\tilde{R_1}$, and $\tilde{R_2}$. These signals serve as additional nonlinear reference signals and their LPS are then concatenated with the original fused feature, and another linear layer is used for feature fusion.

Next, an SARNN module is used to estimate a four-channel enhancement filter, which is then applied on the microphone signal and the three learned reference signals to obtain the enhanced target signal $\hat{S}_1$. Finally, an inverse STFT (iSTFT) is used to obtain the waveform $\hat{s_1}$. More details regarding the feature design and network structure can be found in \cite{yu2022neuralecho}.


\subsection{Loss functions}

In the initial stage of this study, a combination of time-domain scale-invariance signal-to-distortion ratio (SI-SDR) \cite{le2019sdr} loss and frequency-domain mean absolute error (MAE) of spectrum magnitude is used as loss function for model training:
\begin{flalign}
Loss_1 = -\text{SI-SDR}(\hat{s}, s) + \lambda \text{MAE} (|\hat{S}|, |S|)
\end{flalign}
Given that the feedback signals have a strong correlation with the target signal, suppressing them could be difficult. 
To further suppress the leakage introduced due to improperly attenuated playback signals, we propose to include a correlation loss:
\begin{flalign}
Loss_{corr} = [1 - corr(\hat{s}_1, s_1)] + corr(\hat{s}_1- s_1, d_{*})
\end{flalign}
The correlation loss is composed of two terms. The first term evaluates the similarity between the estimated and target signals, while the second term measures the similarity between a playback signal $d_*$ and the residual signal in the estimated target.
The modified loss function we used for model training is: 
\begin{flalign}
Loss_2 = Loss_1 + \beta Loss_{corr}
\end{flalign}
To ensure balance among different losses, we set $\lambda$ and $\beta$ to 10000 and 10, respectively, in our implementation.

%
%
%
%

\section{Experimental setup}

\subsection{Data preparation}

The AISHELL-2 \cite{du2018aishell} and INTERSPEECH 2021 AEC Challenge \cite{cutler2021interspeech} datasets are used for carrying out experiments. 
A total number of 10,000 room impulse response (RIR) sets are generated using the image method \cite{allen1979image}, which incorporate random room characteristics and reverberation times (RT60) range of 0 to 0.6 seconds. Each RIR set consists of 6 RIRs, as shown in Figure~\ref{fig:HybridMeeting}. During data generation, a randomly chosen RIR set is utilized to create near-end speech signals and the corresponding playback signals. System delay is defined as a random value within the range of $[0.1, 0.3]$ second and microphone nonlinear distortions are simulated using a saturation type of nonlinearity with hard clipping and Sigmoidal function \cite{birkett1996nonlinear, lee2015dnn, zhang2018deep}. The microphone signal is generated as a mixture with a randomly chosen signal-to-feedback ratio (SFR) in the range of $[-20, 5]$ dB, and signal-to-noise (SNR) ratio ranging from -10 dB to 30 dB. We created a total of 10,000, 300, and 500 utterances for training, validation, and testing, respectively. The utterances and RIRs used for generating testing data are different from those used in the training and validation data. The model was trained for 60 epochs using a batch size of 20.

\subsection{Method evaluation}

SI-SDR and perceptual evaluation of speech quality (PESQ) \cite{rix2001perceptual} are used as evaluation metrics to show the playback attenuation performance and quality of enhanced speech. A higher value denotes better performance.



\section{Experimental results}


\subsection{Explorations regarding inputs/reference signals}

%

\begin{table}[!t]
\centering
\caption{Explorations regarding inputs/reference signals.}
\label{table:input}
\resizebox{0.4 \textwidth}{!}{
\begin{tabular}{l|ccc|ccc} \specialrule{1.5 pt}{1 pt}{1 pt}  \hline
  $Net_{2-ch}$, $Loss_1$                 & \multicolumn{3}{c|}{SI-SDR (dB)} & \multicolumn{3}{c}{PESQ} \\ \hline
SFR (dB)       &    -10       &     -5     &     0     &   -10     &      -5  &  10      \\  \hline                 
Unprocessed  & -9.49     &   -4.47        &    0.54         &  1.35  &   1.64     &  2.05         \\
$[y_1, x_1]$  &  4.59   &    7.78        &   10.72       &  2.26 &   2.59     &  2.88        \\
$[y_1, x]$   &  3.24    &     6.40      &    9.28      &  2.06   &   2.38     &      2.71      \\
$[y_1, x_{21}]$  & 4.67  &    7.69       &    10.43     &  2.55    &    2.88    &  3.12           \\
$[y_1, x, x_{21}]$ &  5.25  &    8.21       &    10.98   &  2.58     &     2.89   &  3.11         \\
$[y_1, x_{21}, x]$  &  \textbf{5.31}   &    \textbf{ 8.53}      &     \textbf{11.42}      &   \textbf{2.69}     &   \textbf{2.99}     &  \textbf{3.23}       \\
 \hline
\specialrule{1.5 pt}{1 pt}{1 pt}
\end{tabular}
}
\end{table}

To investigate the impact of reference signals on model performance, we conducted experiments with different SFRs, and the results are summarized in Table~\ref{table:input}. 
To diminish the influence of model size difference during the comparison, we remove the deep filtering branches related to $\tilde{R}_2$ in the network and use the resulting simplified network, denoted as  ``$Net_{2-ch}$'', for training the inputs with either 2 or 3 channels. 
This implies that when using an input with three channels, we extract intermediate signals from the first two channels, while the third channel is only used for feature extraction. 
Using the integrated loudspeaker signal $x_1$ as a reference is the most straightforward way to train the model, and the model trained with [$y_1$, $x_1$] is referred to as the ``initial'' model. 
Among the models with 2-channel inputs, the ``initial'' model achieves better SI-SDR in most cases, while its speech quality (PESQ) is not better than that of using $x_{21}$ as the reference signal. 
Models trained using separated reference signals (3-channel inputs) consistently outperformed models using an integrated reference signal $x_1$. For $Net_{2-ch}$, the order of reference signals $x$ and $x_{21}$ determines from which $\tilde{R}_1$ is extracted. The model trained with [$y_1$, $x_{21}$, $x$] as input achieves the best overall performance.


\subsection{Explorations regarding loss functions}

%

\begin{table}[!t]
\centering
\caption{Explorations regarding loss functions.}
\label{table:loss}
\resizebox{0.45 \textwidth}{!}{
\begin{tabular}{ll|ccc|ccc} \specialrule{1.5 pt}{1 pt}{1 pt}  \hline
\multicolumn{2}{l|}{ $Net_{2-ch}$, $[y_1, x_1]$ }                                & \multicolumn{3}{c|}{SI-SDR (dB)} & \multicolumn{3}{c}{PESQ} \\ \hline
SFR (dB)    &   &    -10       &     -5     &     0     &   -10     &      -5  &  10      \\  \hline   
\multicolumn{2}{l|}{Unprocessed}     & -9.49     &   -4.47        &    0.54         &  1.35  &   1.64     &  2.05             \\  \hline   
\multicolumn{2}{l|}{$Loss_{1}$} &  4.59   &    7.78        &   10.72       &  2.26 &   2.59     &  \textbf{2.88}         \\  \hline   
\multicolumn{1}{l|}{\multirow{4}{*}{$Loss_{2}$}} &$d^{s}_{21}$  &3.95  &    6.58       &   8.52       &   2.13    &   2.43     &  2.67          \\
\multicolumn{1}{l|}{}                            &$d_{21}$  & \textbf{4.96} &     \textbf{7.92}      &   \textbf{10.83}       &  \textbf{2.27}   &    \textbf{2.60}    &   \textbf{2.88}          \\
\multicolumn{1}{l|}{}                            & $d^{s}_{21}$+$d^{s}_{11}$ & 4.32 &    6.98       &    9.03      &  2.26   &  2.56      &   2.79         \\
\multicolumn{1}{l|}{}                            & $d_{21}+d_{11}$ & 4.74 &   7.65        &    10.32       &  2.27   &   2.58     &  2.86         \\  \hline \specialrule{1.5 pt}{1 pt}{1 pt} 
\end{tabular}
}
\end{table}


Table~\ref{table:loss} compares the performance of models trained with $Net_{2-ch}$, input [$y_1$, $x_1$],  and $Loss_2$ calculated using different playback $d_*$. 
The results show that incorporating the correlation loss does not consistently lead to performance improvement while using $Loss_2$ calculated based on $d_{21}$ yields the best performance and outperforms the model trained using $Loss_1$, especially in terms of SI-SDR. 
This is because $d_{21}$ is more difficult to suppress due to a lack of direct reference signal, and utilizing it in the calculation of $Loss_2$ helps further attenuating the leakage.

\subsection{Proposed method for joint AEC and AHS}


\begin{table}[!t]
\centering
\caption{Proposed method for feedback suppression.}
\label{table:final}
\resizebox{0.47 \textwidth}{!}{
\begin{tabular}{l|ccc|ccc} \specialrule{1.5 pt}{1 pt}{1 pt}  \hline
Input: $[y_1, x_{21}, x]$    & \multicolumn{3}{c|}{SI-SDR} & \multicolumn{3}{c}{PESQ} \\  \hline
SFR (dB)                              & -10       & -5      & 0      & -10     & -5      & 0     \\ \hline
Unprocessed                     & -9.49     &   -4.47        &    0.54         & 1.35  &   1.64     &  2.05     \\
$Net_{2-ch}$, $Loss_1$ & 5.31  &    8.53      &     11.42     &   2.69      &   2.99     &  3.23           \\
$Net_{3-ch}$, $Loss_1$  & 6.11  &    9.09      &  11.82      & \textbf{2.73}     &   \textbf{3.07}      &   \textbf{3.31}    \\
$Net_{3-ch}$, $Loss_2$ &  \textbf{6.47} &    \textbf{9.24}      &   \textbf{11.87}     &  \textbf{2.73}       &    3.03     &  3.28       \\
 \hline \specialrule{1.5 pt}{1 pt}{1 pt}  
\end{tabular}
}
\end{table}

\begin{figure}[!t]
\centering
     \includegraphics[width=0.99\columnwidth]{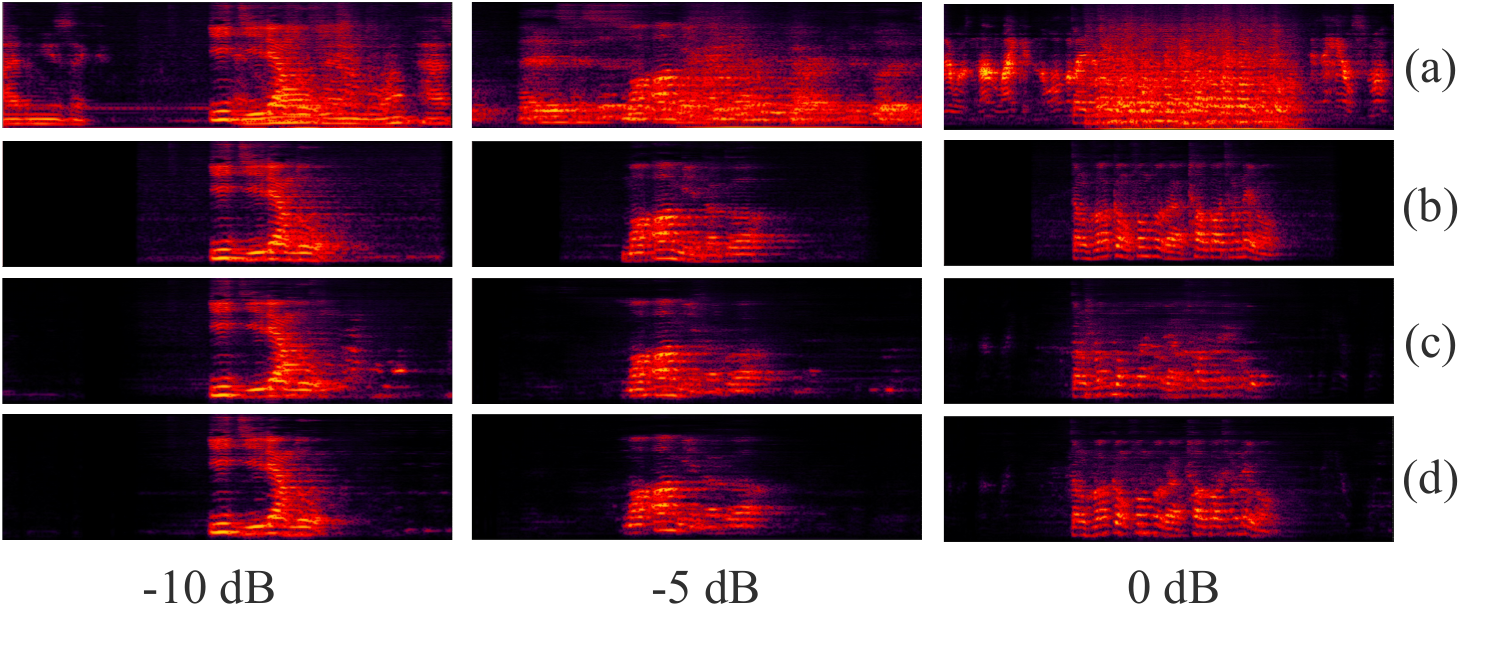}
      \caption{Spectrograms of processed signals obtained under different SFR levels: (a) unprocessed signal, (b) target signal, (c) output of the ``initial'' model, and (d) output of the best model. }
      \label{fig:spectrogram}
\end{figure}

We combine the findings made through explorations regarding inputs and loss functions and train a model to achieve the best feedback suppression. 
Specifically, we utilize a 3-channel input $[y_1, x_{21}, x]$, $Loss_2$ with $d_{21}$, and the network illustrated in Figure~\ref{fig:NeuralEcho}, denoted as ``$Net_{3-ch}$'', for model training. The performance comparison results presented in Table~\ref{table:final} demonstrate that using ``$Net_{3-ch}$'' results in better performance than using ``$Net_{2-ch}$", and employing the modified loss function, $Loss_2$, could further improve playback attenuating performance. 
We also provide spectrograms of processed signals obtained using the ``initial" model and the best-performing model in Figure~\ref{fig:spectrogram} to further illustrate the efficacy of our proposed approach for joint acoustic echo and acoustic howling suppression.

%
%

\section{Conclusion}

We have proposed a deep learning based method for addressing audio-related problems in hybrid meetings. 
Our proposed method treats acoustic echo and acoustic howling as an integrated feedback problem and achieves simultaneous AEC and AHS using a teacher-forcing learning strategy. 
By converting the recursive feedback suppression problem into a speech separation problem, an SARNN model is utilized to extract the target speech from microphone recording with multiple reference signals as additional inputs. 
The impact of input signals, loss functions on joint AEC and AHS performance has been investigated. 
Future work includes considering practical issues such as computational complexity and investigating using cascaded network to suppress acoustic echo and howling gradually.

\newpage

\bibliographystyle{IEEEtran}
\bibliography{JointAECAHS}

\end{document}